# Trip Length Distribution Under Multiplicative Spatial Models of Supply and Demand: Theory and Sensitivity Analysis

Daniele Veneziano[1] and Marta C. González[1,2]

[1] Department of Civil and Environmental Engineering

[2] Engineering Systems Division and Operations Research Center

Massachusetts Institute of Technology

77 Massachusetts Avenue, Cambridge MA 02139.

**Abstract**

We propose new probabilistic models for the spatial distribution of supply and demand and use the models to determine how the trip length distribution is affected by the relative shortage or excess of supply, the spatial clustering of supply and demand, and the degree of attraction or repulsion between supply and demand at different spatial scales. The models have a multiplicative structure and in certain cases possess scale invariance properties. Using detailed population data in metropolitan US regions validates the demand model. The trip length distribution, evaluated under destination choice models of the intervening opportunities type, has quasi-analytic form. We take advantage of this feature to study the sensitivity of the trip length distribution to parameters of the demand, supply and destination choice models. We find that trip length is affected in important but different ways by the spatial density of potential destinations, the dependence among their attractiveness levels, and the correlation between supply and demand at different spatial scales.

## 1. Introduction

Transportation analysts, economists and planners have long been interested in trip forecasting under deterministic or stochastic supply and demand conditions (*Wilson and Kirby*, 1980; *Fernández and Friesz*, 1983). Applications are numerous. For example, origin-destination fluxes and trip length distributions are needed to predict the spreading of global epidemics (*Grenfell*, 2006; *Balcan*, 2009) or develop urban transportation models (*Ben Akiva and Lerman*, 1985, *Sheffi*, 1985; *Ortúzar and Willumsen*, 1990). A first order approach to estimate the number of trips $T_{ij}$ between two locations $i$ and $j$ with populations $N_i$ and $N_j$ is to assume that $T_{ij}$ is some function of distance $r_{ij}$ between the two locations and is proportional to the product $N_iN_j$ (*Casey*, 1955; *Fisk*, 1988). These so-called gravity models work well at large distances, for which the distribution of population and services can be coarse grained (e.g. *Balcan*, 2009). At shorter distances, for example within cities, one must use more detailed models that account for the spatial distribution of supply and demand and the attractiveness of different potential destinations. These more detailed modeling approaches are made possible by the availability of supply and demand data at increasingly fine resolution.

Here we focus on the distribution of trip length when supply and demand have random distributions on the geographical plane with different statistical characteristics and degrees of inter-dependence and destinations are chosen according to an intervening opportunities model (*Schneider*, 1959).It is intuitively clear that trip length must depend on the spatial distribution of supply *S* and demand *D*. Our first objective is to introduce a simple but rich class of random spatial distribution models for *S* and *D*. The models have a multiplicative structure (meaning that the fluctuations of *D* and *S* at different spatial scales combine in a multiplicative rather than additive way) and include an attractive or repulsive dependence between supply and demand that may vary with resolution. For certain choices of the parameters, the models possess a scale-invariance property called multifractality. When it is present, multifractality greatly reduces the number of model parameters, simplifies their inference, and causes the trip length distribution to have a power-law lower tail.

The second objective is to use the proposed models to study how the trip length distribution depends on the overall level, spatial clustering, and degree of dependence between supply and demand at different geographic scales, among other factors. As already noted, spatial dependence between *S* and *D* may be in the form of attraction or repulsion and these opposite tendencies may coexist at different spatial resolutions. For example, at large scales many services tend to follow the distribution of population, whereas at local scales the same services may be preferentially located in non-residential areas.

We assume that destination choices are made according to an intervening opportunities model (*Schneider*, 1959). Specifically, if *S* is the number of supply points (potential destinations) within distance *d* from a trip origin, the probability that none of them is chosen (and therefore the probability that the trip length exceeds *d*) is taken to be

$$P_{>d} = e^{-\lambda S^\alpha} \qquad (1)$$

where $\lambda \geq 0$ and $0 < \alpha \leq 1$ are parameters. The quantity $e^{-\lambda}$ is the probability that any single supply location is not sufficiently attractive as destination. The parameter $\alpha$ controls the dependence among the level of attraction of different supply points. When $\alpha = 1$, the supply points have independent attractiveness levels and the nearest one with attractiveness above a given threshold is chosen. Therefore, $S^\alpha$ is the equivalent number of supply locations with independent attractiveness levels. As $\alpha$ decreases, there is an increasing positive dependence among the levels of attraction of different supply points. In the limit as $\alpha \to 0$, the levels of attraction of all supply points become identical and one either chooses the closest one to the trip origin or finds all potential *S* destinations unsuitable. Without difficulty, $S^\alpha$ in Eq. 1 may be replaced with a different monotonically increasing function $g(S)$ such that $g(S) = S$ for $S = 0, 1$.

Section 2 describes the multiplicative models of supply and demand and Section 3 derives the trip length distribution under Eq. 1. Section 3 also gives more explicit results for three classes of multiplicative models, which we call lognormal, beta and beta-lognormal (this nomenclature is borrowed from turbulence and refers to the distribution of the amplitude of the multiplicative fluctuations of supply and demand at different scales). For these models, the trip length distribution has quasi-analytic form, allowing one to make sensitivity analyses without resorting to Monte Carlo simulation. Section 4 analyzes the population inside 64 x 64 km "metropolitan areas" from different regions of the US to determine the appropriateness of the proposed models for demand and extract realistic parameter ranges. Section 5 makes a sensitivity analysis of the travel distance distribution to the parameters of the demand, supply and destination choice models, constrained by the findings of Section 4. Section 6 summarizes the main conclusions and suggests future research directions.

## 2. Multiplicative Models of Supply and Demand

Consider the spatial distribution of some demand $D$ and corresponding supply $S$ in a geographical region of interest $\Omega_0$. In what follows we take $\Omega_0$ to be the unit square. For any given sub-region $\omega \subset \Omega_0$, $D(\omega)$ is the number of trips per unit time with origin in $\omega$ and $S(\omega)$ is the number of potential destinations in $\omega$. Here we propose a class of models for the bivariate random measure $\underline{X}(\omega) = [D(\omega), S(\omega)]$.

### 2.1 Discrete Cascade Models

We specifically assume that $\underline{X}$ results from a cascade process in which the fluctuations at different spatial scales combine in a multiplicative way [loosely speaking, $D$ and $S$ being non-negative, we assume that the fluctuations of $\ln(D)$ and $\ln(S)$ combine additively]. The simplest measures of this type are the discrete cascades originally proposed for energy dissipation in turbulent flow (*Mandelbrot*, 1974). Their construction is well known; see for example *Mandelbrot* (1974), *Schertzer and Lovejoy* (1989), or *Gupta and Waymire* (1993). To generate a bivariate $[D, S]$ cascade, one starts with uniform measure densities $D_0$ and $S_0$ in $\Omega_0$. Then one progressively partitions $\Omega_0$ into $4^1, 4^2, ..., 4^n, ...$ square tiles of side length $2^{-1}, 2^{-2}, ..., 2^{-n}, ...$ Each time a mother tile $\Omega_{n-1}$ at resolution level $n$-1 is partitioned into 4 daughter tiles $\Omega_n$ at resolution level $n$, the demand and supply densities in the daughter tiles are multiplied by independent realizations of non-negative random factors $W_{D_n}$ and $W_{S_n}$, with mean value 1. **Figure 1** illustrates this hierarchical process for the demand $D$. The random vectors $\underline{W}_n = [W_{D_n}, W_{S_n}]$, $n = 1, 2, ...$ are known as the generators of the cascade. We denote the demand and supply in a generic $n$-tile $\Omega_n$ by $D_n$ and $S_n$ and the associated measure densities by $D'_n = D_n / |\Omega_n|$ and $S'_n = S_n / |\Omega_n|$.

While the generators $\underline{W}_n$ have independent values in different $n$-tiles, their components $W_{D_n}$ and $W_{S_n}$ in a given $n$-tile may be dependent. Moreover, the distribution of $\underline{W}_n$ may vary with the resolution level $n$. These features provide important modeling flexibilities. For example, health care and school services should be available to meet demand within some maximum travel distance $d^*$, while at shorter distances their spatial distribution may differ from that of the demand, requiring local travel. In this case $W_{D_n}$ and $W_{S_n}$ should be highly dependent for $n$ below some resolution level $n^*$ that corresponds to $d^*$ and weakly correlated for larger $n$. In some cases ("not in my backyard, but not too far either"), $W_{D_n}$ and $W_{S_n}$ are negatively correlated for $n > n^*$. Making the distribution of $\underline{W}_n$ vary with $n$ produces supplies and demands with different variability at different spatial scales.

If the generators $\underline{W}_n$ have the same distribution for all $n$, the supply and demand measures have a scale invariance property called multifractality (*Parisi and Frish*, 1985; *Gupta and Waymire*, 1990; *Veneziano*, 1999). Specifically, in the multifractal case the measure densities at resolution levels $n$-1 and $n$ are related as

$$\begin{bmatrix} D'_n \\ S'_n \end{bmatrix} \stackrel{d}{=} \begin{bmatrix} W_D & 0 \\ 0 & W_S \end{bmatrix} \begin{bmatrix} D'_{n-1} \\ S'_{n-1} \end{bmatrix} \qquad (2)$$

where $\stackrel{d}{=}$ denotes equality in distribution of the random vectors on the right and left hand sides. Multifractal cascades have just one generator vector $\underline{W} = [W_D, W_S]$, which determines the scale invariant properties of supply and demand as stated in Eq. 2.

For a visual appreciation of multiplicative models, **Figure 2** compares the spatial distribution of population in a region of approximately $10^6$ square kilometers in the central US (**1999 census**, 1km resolution) with a simulation from a multifractal cascade in which the generator $W_D$ has lognormal distribution. While neither the cascade parameters nor the simulation itself have been fine-tuned to best reflect the features of population, the general resemblance between the two images lends credibility to the proposed models.

To illustrate the flexibility of the model, **Figure 3** shows one-dimensional transects of $D$ and $S$ simulations (in black and red, respectively) at resolution level $n = 8$ for different fluctuation and dependence parameters. The central panel of each row assumes independence between supply and demand, whereas the left panels are for cases with negative dependence (repulsion) and the right panels are for cases with positive dependence (attraction). In the top row, $S$ and $D$ have similar spatial variability characteristics, whereas in the bottom row the variability of $D$ is much lower than that of $S$. In all cases, we have set $D_0 = 2S_0$, mainly to separate the $D$ and $S$ plots. The general resemblance of the plots in different panels is due to the use of a common simulation seed, which makes one better appreciate the effect of varying the model parameters.

Other modeling possibilities, which are not illustrated in **Figure 3**, include varying the type of distribution of the generator $\underline{W}$ (for the simulations in **Figures 2 and 3** the distribution is lognormal) and making that distribution depend on the resolution level $n$. One could also include spatial heterogeneity, meaning that the initial densities $D_0$ and $S_0$ and the distribution of the generator might vary with geographic location, for example to account for topography, proximity to water bodies, land use regulations, and other exogenous factors. In what follows we consider only cases with statistically homogeneous supply and demand.

*2.2 Bare and Dressed Quantities*

In multiplicative models, it is useful to distinguish between bare and dressed quantities; see for example *Schertzer and Lovejoy* (1989). The bare values $D_{b,n}$ and $S_{b,n}$ (hereafter simply denoted by $D_n$ and $S_n$) are obtained by terminating the cascade construction at resolution level $n$, whereas the dressed values $D_{d,n}$ and $S_{d,n}$ are the demand and supply in $\Omega_n$ for a cascade that is developed down to infinitesimal scales. Hence, dressed quantities differ from bare quantities due to their inclusion of sub-grid fluctuations. The bare and dressed demands and supplies at resolution level $n$ are given by

$$D_n = (D_0 4^{-n})\prod_{i=1}^{n} W_{D_i}, \qquad S_n = (S_0 4^{-n})\prod_{i=1}^{n} W_{S_i} \qquad (3)$$
$$D_{d,n} = D_n Z_{D,n}, \qquad S_{d,n} = S_n Z_{S,n}$$

where $4^{-n}$ is the area ratio $|\Omega_n|/|\Omega_0|$ and $Z_{D,n}$ and $Z_{S,n}$ are so-called dressing factors.

In practice, interest is in the dressed quantities, but a dressed analysis is more complicated than a bare one due to the need to find the distribution of the dressing vectors $\underline{Z}_n = [Z_{D,n}, Z_{S,n}]$. Methods to numerically calculate the distribution of the dressing factor $Z$ for scalar multifractal cascades exist (*Veneziano and Furcolo*, 2003) and can be extended to vector and non-multifractal cases, but such an effort is hardly justified here, since we have found that bare and dressed trip length distributions are typically close. Therefore, in what follows we focus on the bare case.

### 3. Trip Length Distribution

As before, the region of interest $\Omega_0$ is the unit square, $\Omega_n$ is a generic cascade tile with side length $d_n = 2^{-n}$, and $D_n$ and $S_n$ are the bare demand and bare supply in $\Omega_n$, in trips originated per unit time and number of potential trip destinations, respectively. $D_0$ and $S_0$ are the total bare demand and total bare supply in $\Omega_0$.

We characterize the distribution of trip length through $P_{>n}$ = probability that a generic trip with origin in $\Omega_0$ has destination outside the region $\Omega_n$ of origin. This is also the probability that a generic trip length exceeds $\sim d_n$, where $\sim$ denotes order of magnitude.

To obtain $P_{>n}$, consider a generic trip with origin in $\Omega_0$ and denote by $[S_{n,trip}, D_{n,trip}]$ the random supply and demand in the sub-region $\Omega_n$ where the trip originates. Then

$$P_{>n} = \int_0^\infty e^{-\lambda s^\alpha} f_{S_{n,trip}}(s) ds \tag{4}$$

where the probability density function (PDF) $f_{S_{n,trip}}$ can be calculated as follows. The PDF of $D_{n,trip}$ is related to the PDF of $D_n$ as

$$f_{D_{n,trip}}(D) \propto D f_{D_n}(D) \tag{5}$$

and the conditional distribution of $[S_{n,trip} | D_{n,trip}]$ is the same as the conditional distribution of $[S_n | D_n]$. Therefore

$$f_{S_{n,trip}}(s) = \int_0^\infty f_{D_{n,trip}}(D) f_{S_n | D_n = D}(s) dD \tag{6}$$

The complexity of calculating $P_{>n}$ through Eqs. 4-6 depends on the distribution of the cascade generators $\underline{W}_i = \{W_{D_i}, W_{S_i}\}, i = 1, 2, \ldots$. Before deriving more explicit results for lognormal, beta, and beta-lognormal cascades, we mention a simplifying approximation that all these models make. Following our definition of supply and demand, $S$ and $D$ are discrete variables with non-negative integer values 0, 1, …, whereas in both Eqs. 4-6 and the models to be described next, $S$ and $D$ are allowed to have any real non-negative value. For example, for $S$ and $D$ discrete the probability density functions in Eqs. 4-6 should be replaced with probability mass functions and the integrals should be replaced with summations. An appropriate discrete representation of supplies and demands would consider the supply $S(\omega)$ and demand $D(\omega)$ in a region $\omega \subset \Omega_0$ as generated by a doubly-stochastic Poisson (DSP) process in which the present models apply to spatially varying intensities $\lambda_S(\omega)$ and $\lambda_D(\omega)$ and the conditional supply and demand $[S(\omega) | \lambda_S(\omega)]$ and $[D(\omega) | \lambda_D(\omega)]$ are Poisson with mean values $\lambda_S(\omega)$ and $\lambda_D(\omega)$, respectively. Equations 4-6 simplify this DSP model by setting $S(\omega) = \lambda_S(\omega)$ and $D(\omega) = \lambda_D(\omega)$. For $\lambda_S(\omega)$ and $\lambda_D(\omega)$ much larger than 1, which is often the case in travel modeling, the effect of this approximation is small.

The main source of distortion of Eq. 4 is the fact that one allows $S_{n,trip}$ to have non-integer values in the range [0, 1]. To reduce these distortions, which are significant only in the lower tail of the trip length distribution when $\alpha \neq 1$, we replace the density $f_{S_{n,trip}}$ in the range [0,1] with a probability mass $P_1 = \int_0^1 s f_{S_{n,trip}}(s) ds$ at $S = 1$ and a probability $P_0 = \int_0^1 (1-s) f_{S_{n,trip}}(s) ds$ at $S = 0$. With this replacement, Eq. 4 becomes

$$P_{>n} = P_0 + e^{-\lambda}P_1 + \int_1^\infty e^{-\lambda s^\alpha} f_{S_{n,trip}}(s)ds \tag{7}$$

*3.1 Lognormal Cascades*

If the log generators $\ln(W_{D_i})$ and $\ln(W_{S_i})$ have joint normal distribution with variances $\sigma^2_{W_{D_i}}$ and $\sigma^2_{W_{S_i}}$, mean values $-\frac{1}{2}\sigma^2_{W_{D_i}}$ and $-\frac{1}{2}\sigma^2_{W_{S_i}}$ and correlation coefficient $\rho_{LNi}$, then $\ln(D_n)$ and $\ln(S_n)$ have joint normal distribution with mean values $m_D$ and $m_S$, variances $\sigma^2_D$ and $\sigma^2_S$ and correlation coefficient $\rho$ (for simplicity, the index $n$ is omitted) given by

$$\sigma^2_D = \sum_{i=1}^n \sigma^2_{W_{D_i}}, \qquad m_D = \ln(D_0 4^{-n}) - \frac{1}{2}\sigma^2_D$$

$$\sigma^2_S = \sum_{i=1}^n \sigma^2_{W_{S_i}}, \qquad m_S = \ln(S_0 4^{-n}) - \frac{1}{2}\sigma^2_S \tag{8}$$

$$\rho = \frac{\sum_{i=1}^n \rho_{LNi}\sigma_{W_{D_i}}\sigma_{W_{S_i}}}{\sigma_D \sigma_S}$$

To obtain the distribution of $\ln(S_{n,trip})$ in Eq. 6, we first calculate the PDF of $D_{n,trip}$ in Eq. 5. After some algebra, one obtains

$$f_{D_{n,trip}}(d) = \frac{1}{d\sqrt{2\pi}\sigma_D} e^{-[d-(m_D+\sigma^2_D)]^2/2\sigma^2_D} \tag{9}$$

meaning that $\ln(D_{n,trip})$ has normal distribution with mean value $m_D + \sigma^2_D = \ln(D_0 4^{-n}) + \frac{1}{2}\sigma^2_D$ and variance $\sigma^2_D$. We also note that the conditional variable $[\ln S_n | D_n]$ has normal distribution with mean value and variance given by

$$m_{\ln S_n | D_n} = m_S + \rho\frac{\sigma_S}{\sigma_D}[\ln(D_n) - m_D]$$

$$\sigma^2_{\ln S_n | D_n} = \sigma^2_S(1-\rho^2) \tag{10}$$

It follows from Eqs. 6, 9 and 10 that

$$\ln(S_{n,trip}) \sim N(m_S + \rho\sigma_S\sigma_D, \sigma^2_S) \tag{11}$$

where $m_S = \ln(S_0 4^{-n}) - \frac{1}{2}\sigma_S^2$ and $N(m,\sigma^2)$ is the normal distribution with mean value $m$ and variance $\sigma^2$. In Appendix A we show that, in the multifractal case, as the resolution level $n \to \infty$, the non-exceedance probability $P_{\leq n} = 1 - P_{>n}$ becomes a power function $P_{\leq n} \propto d_n^\gamma$ of distance $d_n = 2^{-n}$, with exponent

$$\gamma = 2 - \frac{\rho_{LN} \sigma_{W_S} \sigma_{W_D}}{\ln(2)} \tag{12}$$

*3.2 Beta Cascades*

As models of supply and demand, lognormal measures of the type described above have two main limitations: 1. they are unable to represent cases when the supply and demand are positive only in a subset of the region $\Omega_0$, and 2. while lognormal models are able to incorporate positive and negative correlation between supply and demand, they cannot express extreme forms of dependence like mutual exclusion. In some cases, mutual exclusion at local scales is an important characteristic of supply and demand, with significant effects on travel distance. In order to include these features, one may add a "beta component" to the lognormal models as described below (see also *Frisch et al.*, 1978; *Novikov and Stewart,* 1964; and *Schertzer and Lovejoy,* 1989). Beta-lognormal ($\beta$-LN) cascades can be obtained as products of independent purely-LN cascades of the type described in Section 3.1 and purely-$\beta$ cascades, as described here. Combined $\beta$-LN models are dealt with in Section 3.3 (for an application of $\beta$-LN models to rainfall, see *Over and Gupta,* 1996).

The generators $\underline{W}(n) = [W_D(n), W_S(n)]$ of a bivariate $\beta$ cascade have a discrete distribution with probability masses concentrated at four $(w_D, w_S)$ points:

$$\begin{array}{ll} mass\ P_{00}\ at\ (0,0), & mass\ P_{D0}\ at\ (1/P_D, 0), \\ mass\ P_{0S}\ at\ (0, 1/P_S), & mass\ P_{DS}\ at\ (1/P_D, 1/P_S) \end{array} \tag{13}$$

where $P_D = P_{D0} + P_{DS}$, $P_S = P_{0S} + P_{DS}$, and $P_{00} + P_{D0} + P_{0S} + P_{DS} = 1$. Note that $P_D$ and $P_S$ are also the marginal probabilities $P[W_D > 0]$ and $P[W_S > 0]$, respectively. In general, the probabilities in Eq. 13 vary with the cascade level $n$, whereas in the multifractal case they are the same for all $n$. The joint distribution in Eq. 13 has mean values $E[W_D] = E[W_S] = 1$, variances $Var[W_D] = (1 - P_D)/P_D$ and $Var[W_S] = (1 - P_S)/P_S$, and correlation coefficient

$$\rho_\beta = Corr[W_D, W_S] = \frac{P_{DS} - P_D P_S}{\sqrt{P_D P_S (1 - P_D)(1 - P_S)}} \tag{14}$$

where $\rho_\beta$ ranges from $-1$ and $1$. One may parameterize this distribution in terms of $\{P_D, P_S, \rho_\beta\}$ and obtain $P_{DS}$ from Eq. 14. Once $\{P_D, P_S, P_{DS}\}$ are known, the probabilities in Eq. 13 are obtained as

$$P_{D0} = P_D - P_{DS}, \qquad P_{0S} = P_S - P_{DS}, \qquad P_{00} = 1 - P_{D0} - P_{0S} - P_{DS} \tag{15}$$

When they are developed to infinite resolution, beta measures $D$ and $S$ on the plane have fractal support with fractal dimension $2 + \log_2(P_D)$ and $2 + \log_2(P_S)$, respectively. In order for these random measures to be non-degenerate, the fractal dimension must be positive; hence $P_D$ and $P_S$ must satisfy $P_D > 0.25$ and $P_S > 0.25$.

The trip length analysis using bare quantities is simple. The reason is that the demand $D_n$ in Eq. 2 is either 0 or $d_n = D_0 [4^n \prod_{i=1}^n P_D(i)]^{-1}$, the supply $S_n$ is either 0 or $s_n = S_0 [4^n \prod_{i=1}^n P_S(i)]^{-1}$, and the joint distribution of $D_n$ and $S_n$ comprises 4 probability masses:

$$\begin{aligned}
P_{11_n} &= \prod_{i=1}^n P_{DS}(i), & &\text{at } (d_n, s_n) \\
P_{10_n} &= \prod_{i=1}^n P_D(i) - P_{11_n}, & &\text{at } (d_n, 0) \\
P_{01_n} &= \prod_{i=1}^n P_S(i) - P_{11_n}, & &\text{at } (0, s_n) \\
P_{00_n} &= 1 - P_{10_n} - P_{01_n} - P_{11_n}, & &\text{at } (0,0)
\end{aligned} \tag{16}$$

By conditioning on the four possible values of $[D_n, S_n]$ in Eq. 16, the probability $P_{>n}$ that the travel distance exceeds $\sim 2^{-n}$ can be obtained as

$$P_{>n} = \frac{\sum_{i=0}^1 \sum_{j=0}^1 P_{ij_n} D_{ij_n} P_{>n_{ij}}}{\sum_{i=0}^1 \sum_{j=0}^1 P_{ij_n} D_{ij_n}} \tag{17}$$

where the probabilities $P_{ij_n}$ are given in Eq. 16 and the demands $D_{ij_n}$ and probabilities $P_{>n_{ij}}$ are

$$\begin{aligned}
P_{>n_{ij}} &= \begin{cases} 1, & \text{for } j = 0 \\ \exp\{-\lambda s_n^\alpha\}, & \text{for } j = 1 \end{cases} \\
D_{ij_n} &= \begin{cases} 0, & \text{for } i = 0 \\ d_n, & \text{for } i = 1 \end{cases}
\end{aligned} \tag{18}$$

Using Eq. 18, Eq. 17 simplifies to

$$P_{>n} = \frac{P_{10_n} + P_{11_n} e^{-\lambda s_n^\alpha}}{P_{10_n} + P_{11_n}} \qquad (19)$$

### *3.3 Beta-Lognormal Cascades*

Univariate and bivariate $\beta$-LN cascades are obtained as the product of independent lognormal and models. The resulting supply and demand are zero in regions where the beta component is zero and are magnified relative to the lognormal-model values at all other locations. The probability $P_{>n}$ is given by

$$P_{>n} = \frac{\sum_{i=0}^{1}\sum_{j=0}^{1} P_{ij_n} E[D_{ij_n}] P_{>n_{ij}}}{\sum_{i=0}^{1}\sum_{j=0}^{1} P_{ij_{b,n}} E[D_{ij_n}]} = \frac{P_{10_n} + P_{11_n} P_{>n_{11}}}{P_{10_n} + P_{11_n}} \qquad (20)$$

where the probabilities $P_{10_n}$ and $P_{11_n}$ are given in Eq. 16 and $P_{>n_{11}}$ is given by Eq. 7, with the distribution of $S_{n,trip}$ in Eq. 11 and $S_0$ in that equation replaced by $S_0 / \prod_{i=1}^{n} P_S(i)$.

## 4. Analysis of US Population Data

In this section we make a scale-dependent analysis of the detailed US population using Oak Ridge National Lab's LandScan database (**http://www.ornl.gov/sci/landscan/**). The data is obtained from satellite observations and it represents an "ambient population" which is an estimated average over 24 hours. The spatial resolution is 2400 times more refined than the previous standard, and it consists of information on population density at 1 km resolution (30" X 30"). Since the demand $D$ is often directly related to population density, this analysis is used to show whether demand models of the type described in Section 3 are realistic and to produce plausible parameter ranges. Section 4.1 discusses how one can determine from data whether a random measure $D$ has multifractal properties and how to estimate the parameters of the $\beta$-LN model in Section 3, in the multifractal and non-multifractal cases. Special cases of this procedure apply to purely lognormal and purely beta cascades. Section 4.2 applies these analysis tools to population in US metropolitan regions with different population density.

### *4.1 Inference of Multifractality and Multiplicative Parameters for D*

A hallmark of multifractality is that the moments of the measure density $D'_n = D_n / |\Omega_n|$ are power functions of the resolution $2^n$,

$$E[D_n'^q] \propto 2^{nK(q)} \qquad (21)$$

where $K(q) = \log_2(E[W_D^q])$ is a concave function. Hence, multifractality holds if the log-log plots of the moments against resolution are linear and $K(q)$ is the slope of those linear plots. For $\beta$-LN cascades, $K(q)$ is a quadratic function,

$$K(q) = -(\log_2 P_D)(q-1) + \frac{V_D}{2}(q^2 - q) \qquad (22)$$

with beta parameter $P_D$ and lognormal parameter $V_D = \sigma^2_{W_D}/\ln(2)$ (in multifractal analysis, $V_D$ is a more frequently used parameter than $\sigma^2_{W_D} = Var[\ln(W_D)]$). There are different ways to estimate these parameters using Eq. 22. A simple one is to use the empirical values of $K(0)$ and $K(2)$. Then

$$\begin{aligned} P_D &= 2^{K(0)} \\ V_D &= K(2) + K(0) \end{aligned} \qquad (23)$$

If $K(0) = 0$, then $P_D = 1$ and the cascade is purely lognormal, and if $K(2) = -K(0)$, then $V_D = 0$ and the cascade is purely beta.

If the moments do not scale with resolution, one may use the local moment slopes at resolution level $n = 1, 2, \ldots$ to estimate $K_n(q)$ and obtain resolution-dependent parameters $P_{D_n}$ and $V_{D_n}$ by inserting $K_n(0)$ and $K_n(2)$ into Eq. 23.

The above can be extended to bivariate measures $\underline{X} = [D,S]$, but for the analysis of $D$ that follows, the scalar theory suffices.

### *4.2 Scaling Analysis of the US Population*

The analyzed population database from LandScan contains the coordinates and population for 3,237,548 points from the continental US. We map the points in a grid with degree resolution $\Delta\theta$=0.00833, the area (a) of each point is corrected according to the change in latitude, where a=hxw, the height (w) is constant and equal to 1km and the width changes as w=$\Delta\theta$RCos(Lat). First we cover most of the continental US with three 2048x2048 km regions, which we call Eastern (E), Central (C), and Western (W) regions; see **Figure 4a**. In analyzing the population data, we focus on what we call metropolitan areas, defined as 64x64 km square regions with relatively high population. After partitioning each of the three regions into 1024 64x64 km sub-regions, we select the sub-regions in a given relative population range (top 5%, top 5-10%, top 10-15%, etc.). Finally, we normalize the population in each selected sub-region to average to 1 (to give equal weight to all the sub-regions in the moment-scaling analysis that follows). **Figure 4b** shows the sub-regions with top 5% population, with color-coding of the 1 km population density after normalization.

For the three large regions, **Figures 5a-c** show log-log (base-2) moment plots against resolution for the metropolitan sub-regions of the E, C and W US in the top 5% population range. Interestingly, the plots show a bi-linear behavior, with different scaling characteristics above and below 16 km. The $K(q)$ functions obtained from the slopes of the moment plots in the two distance ranges are shown in **Figures 5d-f.** Also shown in these figures are parabolic fits to the empirical $K(q)$ functions. The fact that the empirical $K(q)$ values conform well to parabolic functions confirms the validity of the $\beta$-LN multiplicative model.

The analysis of sub-regions in different population ranges produces similar results, in all cases with a scaling break at about 16 km. The $P_D$ and $V_D$ parameters obtained from Eq. 23 in the two scaling regimes and for sub-regions in different areas (E, C and W) and with different population densities are listed in **Table 1**. At scales larger than 16 km, the values of $P_D$ are close to 1, indicating complete population coverage. At smaller scales the values of $P_D$ are lower, especially in the Western US, and tend to decrease with decreasing population. This makes sense, as in less populated regions one would expect the population support to have a lower fractal dimension. The parameter $V_D$ generally increases with decreasing population density, indicating that the multiplicative fluctuations of population density are larger in less populated regions. $V_D$ is much smaller at local scales than at scales larger than 16 km.

The above analysis supports the beta-lognormal multiplicative model for demand and shows multifractal scaling with different beta and lognormal parameters above and below 16 km. The sensitivity study that follows builds on these results.

## 5. Sensitivity Analysis of the Trip Length Distribution

In the $\beta$-LN case, the supply, demand and choice models have the following overall parameterization:

$$
\begin{array}{ll}
S_0, & \text{for the general level of supply in } \Omega_0 \\
V_D, V_S, \rho_{LN}, & \text{for LN scaling} \\
P_D, P_S, \rho_\beta, & \text{for } \beta \text{ scaling} \\
\lambda, \alpha, & \text{for the choice model}
\end{array}
\quad (24)
$$

where $V_D = \sigma^2_{W_D}/\ln(2)$ and $V_S = \sigma^2_{W_S}/\ln(2)$. If the cascade is non-multifractal, the beta and lognormal parameters further vary with the resolution level $n$. Note that the average density $D_0$ affects multiplicatively the number of trips but under the present destination choice model has no influence on the trip length (this is why the parameter $D_0$ is not listed in Eq. 24). Also, when $D$ and $S$ are independent ($\rho_{LN} = \rho_\beta = 0$), the distribution of $D$ is immaterial and the only parameters that matter for the trip length distribution are $S_0, V_S, P_S, \lambda$ and $\alpha$. Making a comprehensive analysis of how the trip length distribution varies with all these parameters would be tedious. Therefore, in what follows we limit ourselves to illustrating a few interesting effects.

We start by considering multifractal cascades of the lognormal type (by setting $P_D = P_S = 1$ and assuming that the lognormal parameters do not vary with $n$) and choose the following base-case parameter values: $V_D = V_S = 1.0$, $\rho_{LN} = 0$, $\lambda = 0.5$, $S_0 = 20$, and $\alpha = 1$ (note that for $\alpha = 1$, the parameters $\lambda$ and $S_0$ affect the trip length distribution through their product $\lambda S_0$, which is the expected number of attractive destinations inside the unit square $\Omega_0$). Then we make sensitivity analyses with respect to $\lambda S_0$, $\alpha$, and $(V_D, V_S, \rho_{LN})$. The reason why we have varied simultaneously the last 3 parameters is that these parameters interact in a non-trivial way. The results are shown in **Figure 6**. All the cumulative distribution plots are on log-log paper, so that a linear behavior for short trip lengths corresponds to a power-law lower tail of the distribution. The asymptotic slope of these lines is given by $\gamma$ in Eq. 12.

As **Figure 6a** shows, changing $\lambda S_0$ has the effect of translating the distribution of travel distance (all other parameters being fixed, the travel distance is proportional to $\sqrt{\lambda S_0}$). Changing $\alpha$ (**Figure 6b**) has a very different effect, which is confined to the body and upper tail of the distribution. As was explained following Eq. 1, under $\alpha = 1$ the potential destinations have independent attractiveness levels. As $\alpha$ decreases, the attractiveness levels are positively correlated and in the limit as $\alpha \to 0$, the distribution of travel distance depends exclusively on the distance $d_{min}$ to the closest potential destination (the other destinations do not matter). In **Figure 6b**, $d_{min}$ is about 0.2. If the nearest potential destination is unattractive, then for $\alpha$ close to 0 all other destinations are also unattractive and the trip length must exceed the unit linear size of the region $\Omega_0$; see black line in **Figure 6b**. The other panels of **Figure 6** show the sensitivity of travel distance to the correlation $\rho_{LN}$ between the spatial fluctuations of supply and demand at any given scale. The sensitivity depends on the amplitude of the fluctuations, which is controlled by the lognormal parameters $V_D$ and $V_S$. For $V_D \to 0$ or $V_S \to 0$, one of the fluctuations is nil and $\rho_{LN}$ has no effect on travel distance. When both $V_D$ and $V_S$ are large, as is the case in **Figure 6c**, the travel distance distribution is very sensitive to the correlation. This can be intuitively appreciated by looking at the simulations in **Figure 3**, which were made under the same positive and negative values of $\rho_{LN}$ as those used in **Figure 6**.

**Figure 7** shows the effects of lack of multifractality by varying the parameters of the lognormal cascade with the resolution level $n$. In **Figure 7a** the correlation parameter $\rho_{LN}$ is set to the very high value of 0.99 and what varies is the parameter $V_D = V_S = V$. The top and bottom plots are reference multifractal cases with $V$ constant and equal to either 1 (very high spatial variability and statistically short trips due to the high correlation between supply and demand) or 0.01 (almost uniform distribution of supply and demand inside $\Omega_0$ and statistically longer trips). The other plots are for cases in which there is high spatial variability at large scales and various degrees of variability at smaller scales. The asymptotic power-law exponent at small scales (and hence the slope of the plots for small $d$) depends exclusively on the small-scale parameters.

In **Figure 7b**, what varies is the correlation coefficient $\rho_{LN}$. The bounding plots use $\rho_{LN} = -0.99$ and $0.99$ at all scales, whereas the other plots fix $\rho_{LN} = 0.99$ at large scales and use different correlation values at smaller scales. The highest contrast in $\rho_{LN}$ corresponds to high similarity in the spatial distributions of supply and demand at large scales and strong repulsion at small scales.

**Figure 8** shows the effect of including a beta component to the fluctuations of *D* and *S*. In **Figure 8a**, the beta parameters $P_D$ and $P_S$ are kept constant ($P_D = 0.95$ is close to the values obtained for the US population - see **Table 1** - and $P_S = 0.50$ reflects the often greater clustering in the concentration of services). The correlation coefficient $\rho_\beta$ varies in different plots. The reason for the rather modest effect of changing $\rho_\beta$ is that, as $P_D \to 1$, the trip length distribution becomes independent of $\rho_\beta$.

**Figure 8b** sets ($P_D = 0.95$, $\rho_\beta = 0.99$) and varies the amount of beta component in the supply. The case with $P_S = 1.0$ corresponds to a purely lognormal model for the supply and produces statistically short trips. As $P_S$ decreases, the spatial clustering of the supply locations increases and the trip length increases. Note that $P_S = 0.25$ produces the smallest possible fractal dimension of the support of *S*, virtually concentrating all the supply at a single spatial location. Finally, **Figure 8c** varies $P_D$ (in a narrow range consistent with **Table 1**) while keeping $P_S$ and $\rho_\beta$ constant… Due to the assumed high correlation between supply and demand fluctuations ($\rho_\beta = 0.99$), the trip length decreases as the demand becomes more spatially clustered ($P_D = 0.9$).

We have also examined the sensitivity of the trip length distribution to the supply model when the demand parameters are taken from **Table 1**. **Figure 9** shows some such results for metropolitan areas in the top 5% population bracket. Unlike the previous figures, we now plot the exceedance probability $P_{>d}$ and use arithmetic scales for the travel distance *d* and the probability $P_{>d}$. The region $\Omega_0$ has size 64 x 64 km, like the metropolitan areas in **Figure 4**. Base-case values for the supply and correlation parameters are **….** **Figure 9a** shows base-case results for areas in the Eastern, Central and Western portions of the US **… explain the differences**… The other panels in **Figure 9** show the sensitivity to the lognormal and beta correlation coefficients ($\rho_{LN}, \rho_\beta$), the lognormal variability of the supply ($V_S$), and the parameter $\lambda S_0$, which have the rough meaning of expected number of attractive supply points in the whole region….**Comment…**

Although **Figures 6-9** explore only some of the sensitivities of the model, the results reveal a rich modeling environment, with qualitatively intuitive but often quantitatively complex effects of the model parameters on the trip length distribution.

## 6. Conclusions

The main contributions of the present study are the introduction of multiplicative cascade models of supply and demand and the derivation of the trip length distribution for the so-called beta-lognormal class of such models, under destination choice models of the intervening opportunities type. To verify the reasonableness of multiplicative models, we have analyzed the spatial distribution of population in metropolitan regions of the US and found that, in good approximation, population has such a multiplicative representation. Taking advantage of the semi-analytical form of the trip length distribution, we have made sensitivity analyses of the trip length distribution with respect to a number of model parameters. The results show interesting dependences of the trip length on the overall level of supply, the spatial variability of supply and demand, the dependence among the attractiveness levels of different potential destination points, the strength of the beta component (which produces regions of various sizes with complete absence of supply or demand), and the degree of dependence between the multiplicative fluctuations of supply and demand at different spatial scales. Results of this type should be of use to transportation modelers and regional planners.

The proposed models could be improved or extended in several ways:

1. Cascade models of the beta-lognormal type are flexible, but are not the only ones possible. For example, the lognormal component could be replaced with a more general log-stable component; see for example Schertzer and Lovejoy, 1987. Log-stable models have an additional parameter (the index of stability $\alpha$, with values in (0,2] and $\alpha = 2$ in the lognormal case). The main difficulty with using log-stable models is that, with few exceptions, stable distributions do not have analytic probability density functions;
2. As was noted in Section 3, one could use a doubly-stochastic Poisson (DSP) representation of the supply $S$ and demand $D$, in which the cascade models of Section 3 apply to the Poisson intensities rather than directly to $S$ and $D$. The effect of using more satisfactory models of the DSP type could be significant in regions and at scales for which the expected value of $S$ falls below 1;
3. In modeling $S$ and $D$, one could explicitly recognize the effect of exogenous factors like topography, proximity to the sea, and land use regulations. This would produce spatially non-homogeneous distributions of $S$ and $D$;
4. We have validated the multiplicative models by analyzing the spatial distribution of population (a surrogate for demand) inside metropolitan areas in the US. One could make more extensive analyses of population and supply data to support the models of $S$ and $D$.
5. Location aware devices like mobile phones are becoming ubiquitous worldwide (*The Economist*, 2010). Studying trip length distributions from mobile phone users (*González*, 2008) can provide further calibration to the theory proposed here.
6. The present study uses destination choice models of the intervening opportunities type and assumes that each potential destination has unlimited supply capacity (hence the level of service and attractiveness of a potential destination point does not depend on how many individuals choose that destination). It would be interesting to compare the trip length distribution under alternative destination choice models (gravity, logistic, minimum total travel distance…), including models in which supply locations have finite serving capacity;

7. In Section 2.2, we draw a distinction between bare and dressed demands and supplies. For simplicity, we have analyzed the trip length using bare quantities, but in reality one is interested in the dressed quantities. An exact dressed analysis poses significant technical difficulties, but it is possible to make accurate approximations to the distributions of the dressing factors in Eq. 3 under which the semi-analytic simplicity of the bare results is retained;
8. It would be interesting to produce extensive sensitivity results for the mean trip length (rather than the entire trip length distribution). The mean trip length is of practical interest because, when multiplied by the total demand in a region, it gives the total expected travel (say, in miles per day) with origin in that region.

## Appendix A: Derivation of Eq. 12

Under multifractality, the variances and correlation coefficient in Eq. 8 are simply

$$\sigma_D^2 = n\sigma_{W_D}^2, \qquad \sigma_S^2 = n\sigma_{W_S}^2, \qquad \rho = \rho_W \tag{A1}$$

and from Eqs. 7 and 11,

$$P_{\leq n} = 1 - P_0 - e^{-\lambda}P_1 - \int_1^\infty e^{-\lambda s^\alpha} f_{S_{n,trip}}(s)ds \tag{A2}$$

One can show that, as $n \to \infty$, $P[S_{n,trip} < s_o] \to 1$ for any positive $s_o$. Therefore, as $n \to \infty$ the integral in Eq. A2 becomes negligible, $P_1 \to E[S_{n,trip}] = e^{m_S + \sigma_S^2/2}$ where $m_S = \ln(S_0 2^{-2n}) - \frac{1}{2}\sigma_S^2 + \rho\sigma_S\sigma_D$, and $1 - P_0 \to P_1$. Then, under $n \to \infty$, Eq. A2 becomes

$$\begin{aligned} P_{\leq n} &= (1 - e^{-\lambda})P_1 \\ &= (1 - e^{-\lambda})S_0 2^{-2n} e^{\rho n \sigma_{W_S} \sigma_{W_D}} \\ &= (1 - e^{-\lambda})S_0 2^{-n[2 - \rho\sigma_{W_S}\sigma_{W_D}/\ln(2)]} \end{aligned} \tag{A3}$$

Equation A3 shows that, at small scales, $P_{\leq n}$ has a power-law dependence on distance $d_n = 2^{-n}$, with exponent in Eq. 12 (one can further show that the exponent is always positive, as it should be).

**Figure Captions**

Figure 1. Illustration of the discrete cascade construction.

Figure 2. Comparison between (a) the population distribution in a 1024x1024 km region in the Central US and (b) a simulation from a lognormal multifractal cascade.

Figure 3. Linear transects of supply and demand generated by lognormal multifractal cascades with different spatial variability parameters $\sigma_D^2$ and $\sigma_S^2$ and correlation coefficients $\rho_{LN}$.

Figure 4. Distribution of population in the Continental US at 1 km resolution: (a) West, Central and East regions and (b) metropolitan 64x64 km areas with top 5% population.

Figure 5. Moment scaling analysis for population in metropolitan US areas with highest 5% population.

Figure 6. Sensitivity of the trip length distribution to various model parameters. Non-indicated parameters are set to their base-case values $V_D = V_S = 1.0$, $\rho_{LN} = 0$, $\lambda = 0.5$, $S_0 = 20$, and $\alpha = 1$.

Figure 7. Sensitivity of the trip length distribution to scale-dependence of the parameters $V_S = V_D = V$ and $\rho_{LN}$.

Figure 8. Sensitivity of trip length distribution to the parameters $P_D, P_S$ and $\rho_\beta$ of the beta fluctuation component.

Figure 9. **Dependence of the exceedance probability $P_{>d}$ on the supply parameters when the demand parameters are fixed to the values in Table 1 for metropolitan areas with top 5% population.**

*Table* 1: Multifractal parameters $P_D$ and $V_D$ for 64 x 64 km metropolitan regions with different population densities and in different parts of the US. The parameters are different at scales smaller or larger than 16 km.

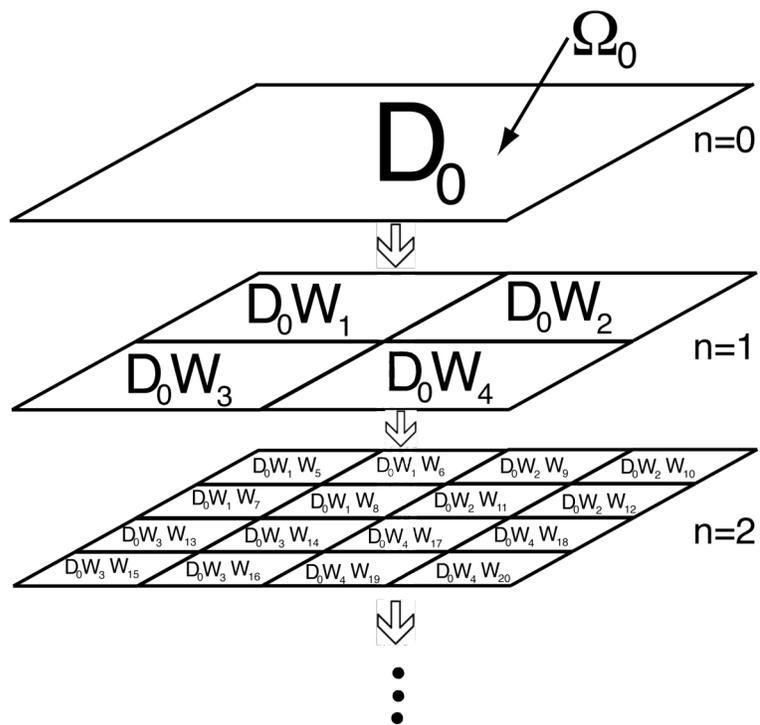

*Figure 1.*

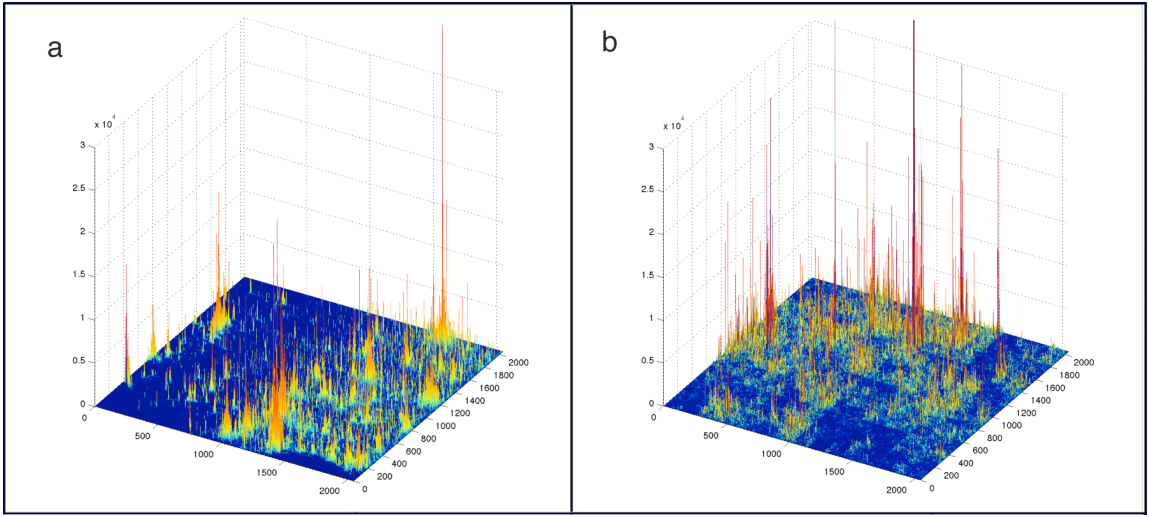

*Figure 2.*

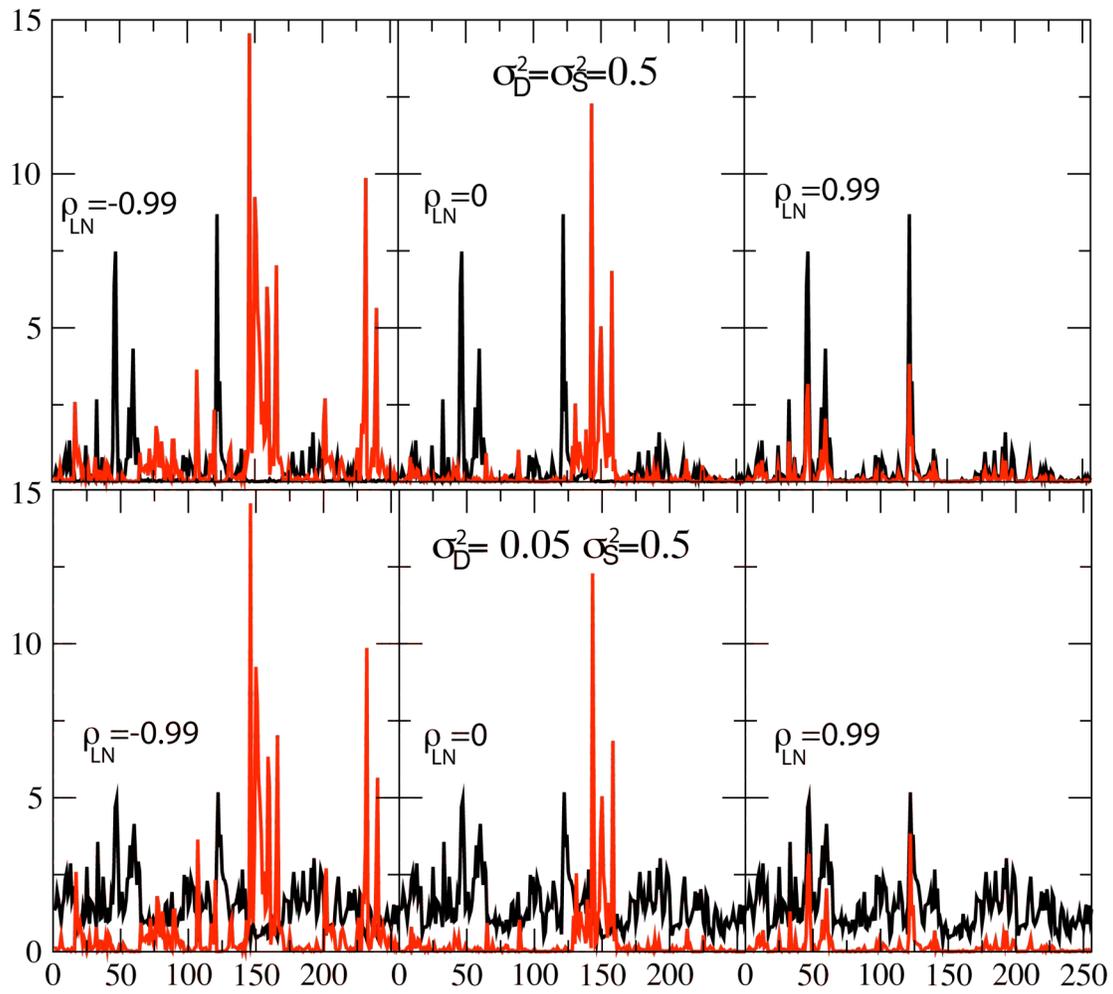

*Figure 3.*

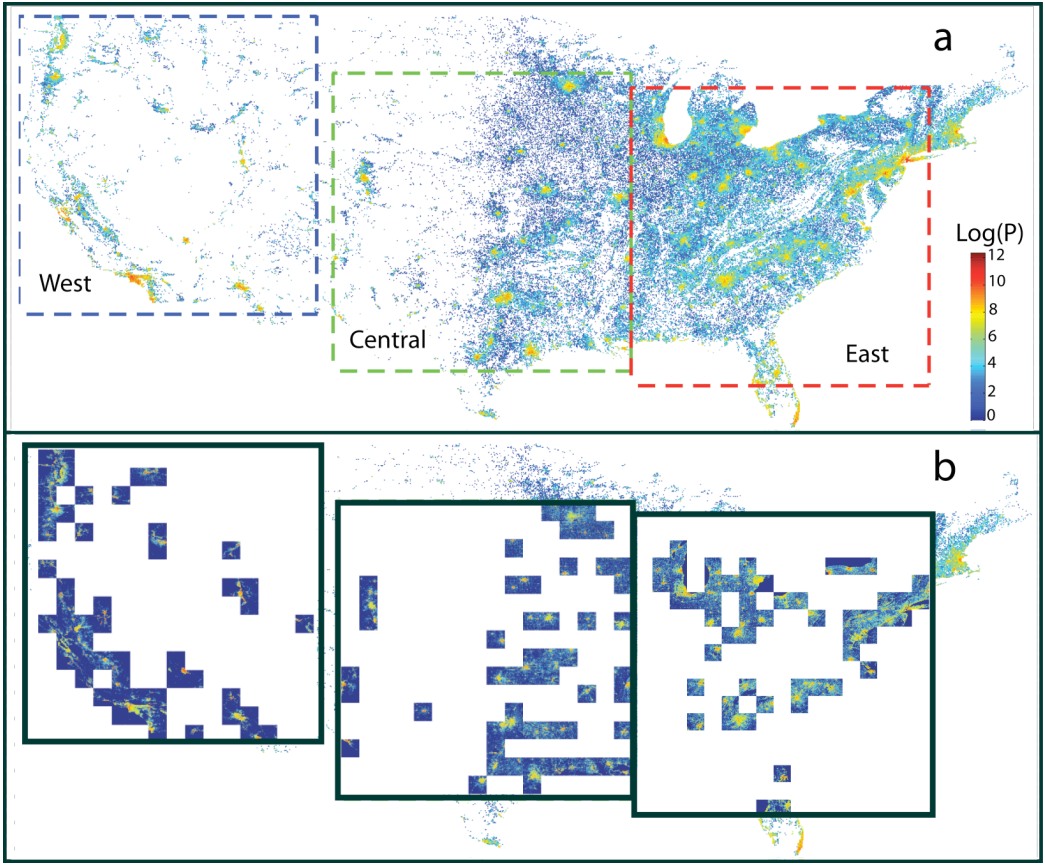

*Figure 4.*

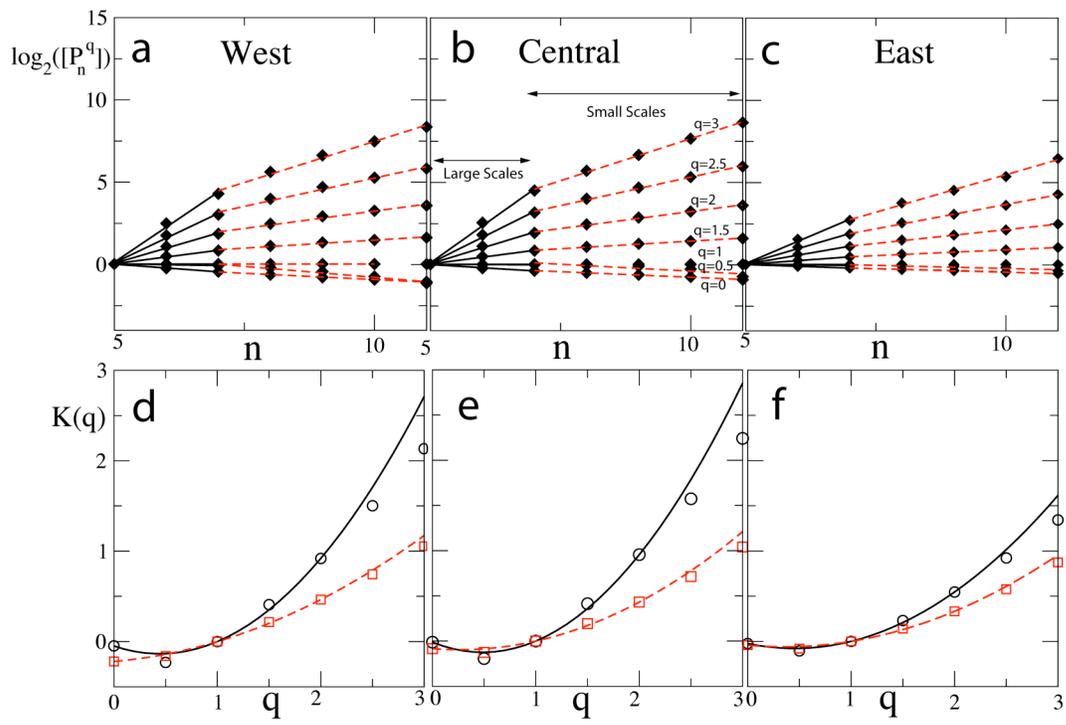

*Figure 5.*

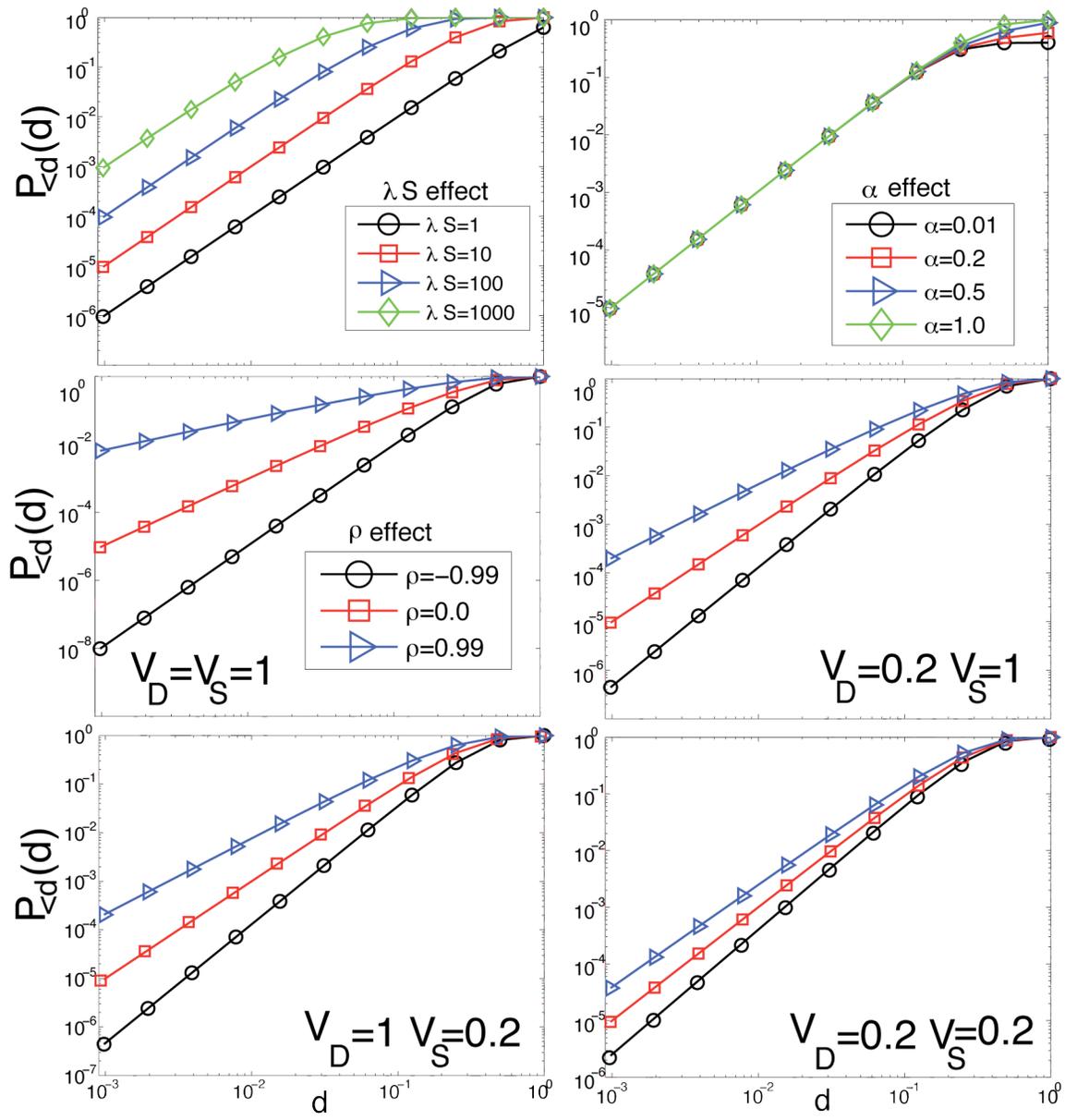

*Figure 6.*

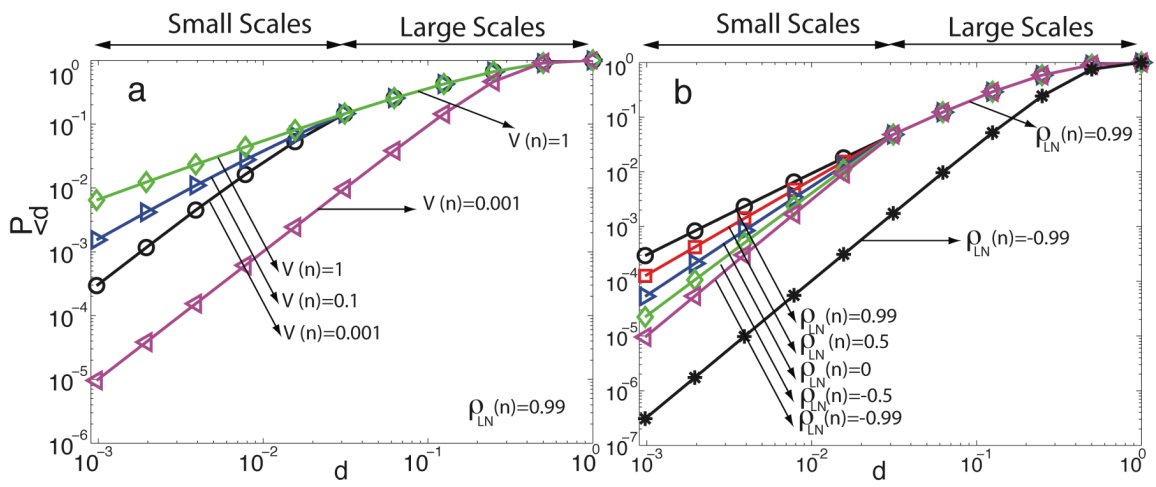

*Figure 7.*

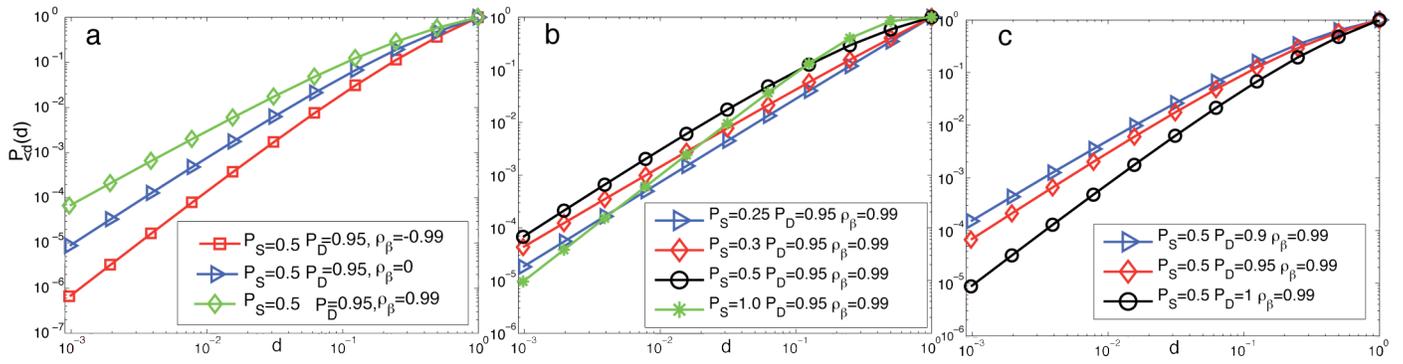

*Figure 8.*

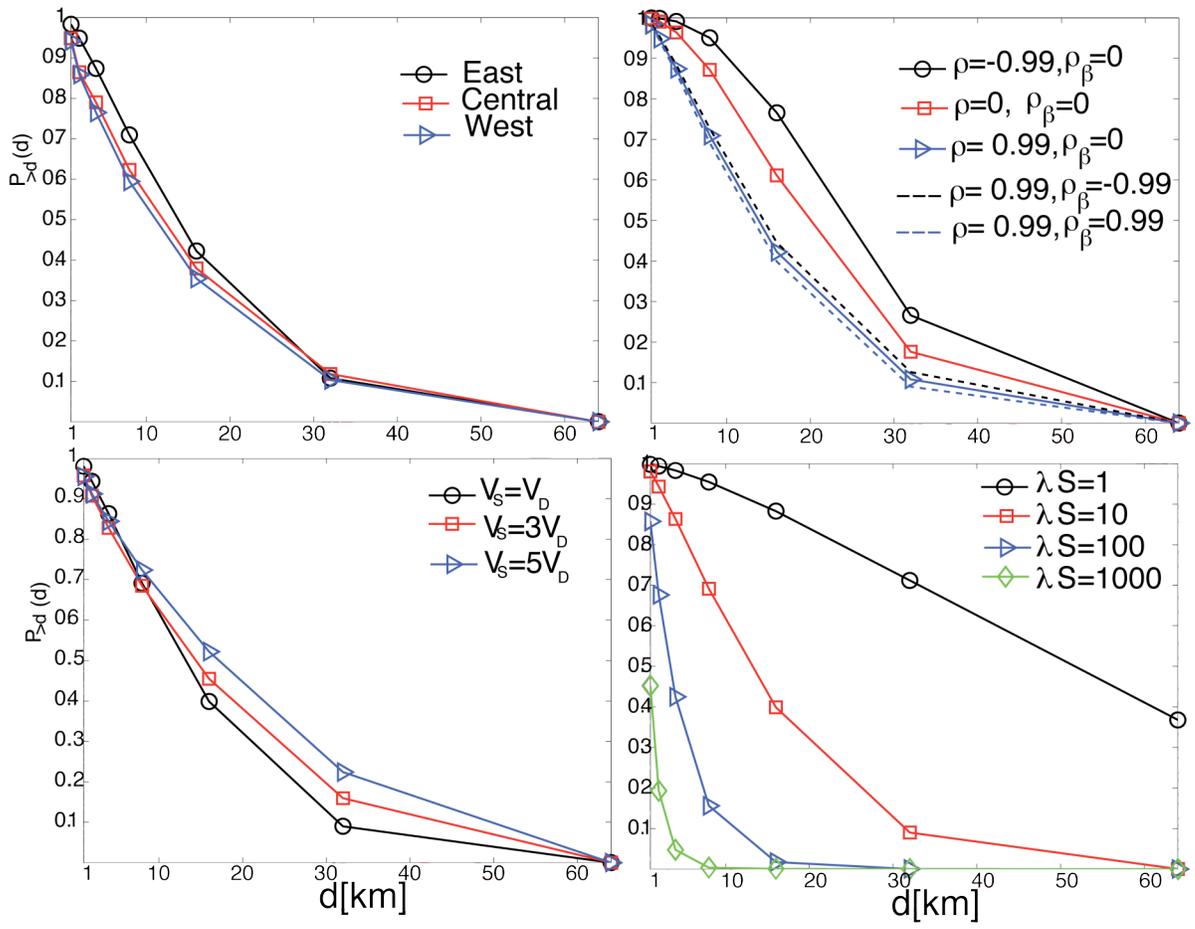

Figure 9.

| $P_D$ | | Top 5% | 5 – 10% | 10 – 15% | 15 – 20% |
|---|---|---|---|---|---|
| East | < 16 km | 0.97 | 0.97 | 0.95 | 0.95 |
| | ≥ 16 km | 0.98 | 0.99 | 0.98 | 0.98 |
| Central | < 16 km | 0.94 | 0.91 | 0.91 | 0.91 |
| | ≥ 16km | 0.99 | 0.99 | 0.99 | 1.00 |
| West | < 16 km | 0.86 | 0.77 | 0.71 | 0.69 |
| | ≥ 16 km | 0.97 | 0.95 | 0.93 | 0.93 |

| $V_D$ | | Top 5% | 5 – 10% | 10 – 15% | 15 – 20% |
|---|---|---|---|---|---|
| East | < 16 km | 0.29 | 0.43 | 0.51 | 0.62 |
| | ≥ 16 km | 0.52 | 0.69 | 0.79 | 0.87 |
| Central | < 16 km | 0.35 | 0.57 | 0.64 | 0.83 |
| | ≥ 16km | 0.94 | 1.01 | 1.14 | 0.93 |
| West | < 16 km | 0.24 | 0.39 | 0.56 | 0.55 |
| | ≥ 16 km | 0.87 | 1.19 | 1.19 | 1.87 |

Table 1.